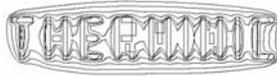

# CHARACTERIZATION OF THERMAL INTERFACE MATERIALS TO SUPPORT THERMAL SIMULATION

*Ralph Schacht[1], Daniel May[1], Bernhard Wunderle[1], Olaf Wittler[2], Astrid Gollhardt[1], Bernd Michel[1], Herbert Reichl[2]*

[1]Fraunhofer Institut für Zuverlässigkeit und Mikrointegration,
[2]Technische Universität Berlin, Forschungsschwerpunkt Technologien der Mikroperipherik,
Gustav-Meyer Allee 25, 13355 Berlin, Germany
ralph.schacht@izm.fraunhofer.de

**ABSTRACT**

In this paper new characterization equipment for thermal interface materials is presented. Thermal management of electronic products relies on the effective dissipation of heat. This can be achieved by the optimization of the system design with the help of simulation methods. The precision of these models relies also on the used material data. For the determination of this data an experimental set-up for a static measurement is presented, which evaluates thermal conductivity and interface resistance of thermal interface materials (e.g. adhesive, solder, pads, or pastes). A qualitative structure-property correlation is proposed taking into account particle size, filler content and void formation at the interface based on high resolution FIB imaging.

The paper gives an overview over the set-up and the measurement technique and discusses experimental and simulation results.

## 1. INTRODUCTION

It is common practice for high power applications to assemble power devices as bare dies in chip on board technique or to use flip chip assemblies with a directly mounted copper or aluminum heat sink. The thermal resistance ($R_{th}$) of the thermal interface material (TIM) is the bottleneck of the thermal heat flow from the active device junction to the cooler. All the more important is the fact of strongly localized hot spots, as heat spreading requires better thermal interface materials as just heat-transfer [1, 2]. If possible, it is useful and common to reduce the thickness of the thermal interface material. But for thinner thermal interface materials the thermal interface resistance (between the silicon and the thermal interface material ($R_{th0,Si-TIM}$) as well as between the thermal interface material and the cooler material ($R_{th0,CM-TIM}$)) is no longer negligible [3, 4]. Therefore it is essential to account for this effect in the measurement set-up.

In the system design process thermal simulation is used among other tools to select the thermal interface material to find the optimum situation.

To build up the simulation model and get a suitable solution it is necessary to know the accurate geometry and material parameters. As , the trend is towards thinner thermal interface materials, the thermal interface resistances should be considered in the thermal simulation model.

There are several standardized test methods (ASTM E 1225-99, ASTM E 1461-01, ASTM E 1530-99, DIN V 54462) as well as in-house-built and market going set-ups to determine the thermal resistance [5].

The most widely used methods are to place the thermal interface material between two plates, heating the top plate and cooling the bottom plate [6, 7, 8] or to use photo or laser flash set-ups [9].

In the paper different thermal interface materials (adhesives) are characterized varying different measurement and material parameters. Thus the thermal conductivity of the materials and the interface resistance is characterized.

It is shown that the interface material can become a key issue for the thermal management of dynamically loaded devices. On this example a model with statically determined material data is validated and discussed in simulation and experiment.

## 2. THERMAL INTERFACE MATERIAL

Thermal Interface Materials as paste, pads , gels, adhesives or solders are used to fulfill the gap or tilt between to surfaces to optimize the thermal path between them. In microelectronics applications the adjacent materials are usually silicon and the heat sink materials Al or Cu. Figure 1 shows on the left hand side a typical power assembly application e.g. for a reverse-side cooled power device in flip-chip technology.

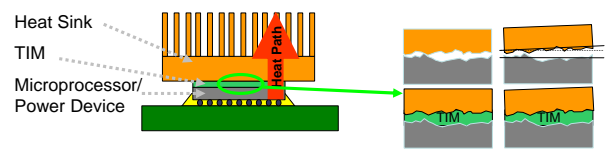

Fig. 1: Thermal interface material is used to fulfill the gap or tilt between to surfaces. For example, on the left a typical reverse-side cooling power assembly application is given.

As mentioned above thinner TIM layers are aimed to minimize the thermal resistance. It is known that the thermal resistance cannot be infinitely decreased by making it thinner. The observed behavior is known as a boundary effect in the interface between e.g. substrate and TIM as well as between TIM and heat sink material. This thermal interface resistances $R_{th0,i}$ has to be added to the bulk resistance of the TIM $R_{th,TIM}$ and leads to an effective thermal resistance $R_{th,eff}$.

$$R_{th,eff} = R_{th0,1} + R_{th,TIM} + R_{th0,2} \quad (Eq.\ 1)$$

## 3. TEST SET-UP
### 3.1. Structure

The idea to develop a new kind of set-up was to realize test conditions as they occur in real assemblies. That means using the same surface materials and assembly technology as in the real device instead of pellets made of adhesives (or solder).

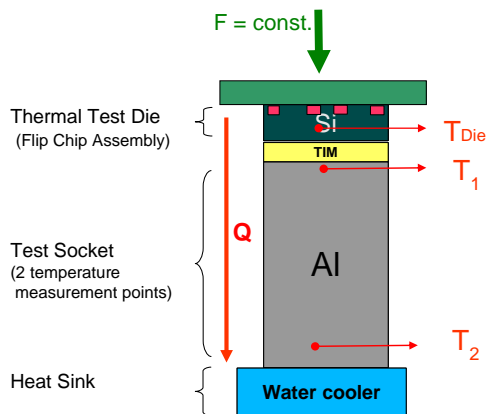

Fig. 2: Principle schematic for new test set-up

Figure 2 shows the schematic set-up for the new method. As in microelectronic assemblies usually silicon surface are assembled on aluminum or copper heat sink material a thermal test die for heating and temperature measuring is used [10].

Therefore a thermal test die was chosen and assembled in flip-chip technology. Using flip-chip technology, it has - to be assured that the die surface is flat, what is especially important for grease characterization. This can be achieved under certain circumstances, as due to thermal mismatch a flip-chip is always slightly curved.

The TIM is instead of a direct mounting on the cold plate, mounted on an test socket (Al or Cu). The test socket is at least placed on the heat sink. The advantage using a test socket is that the heat flow through the TIM could be measured. Another advantage is that the socket is not fixed on the heat sink, so specimens can be easily changed. And while the socket is made out of an inexpensive, market going, low tolerance material which can be used as reference resistor. Due to its being inexpensive, it need not be reused and can be used for interface charac-terization by destructive analysis methods like cross sections or FIB-milling. Here also the exact thickness can be determined. It therefore has the versatility to be used for characterization of adhesive or solder TIMs without damaging the test-set-up.

As in figure 3 is shown for this application a 3x3 thermal test-die chip is used to supply the heat and to measure the chip temperature. The thermal test chip is mounted in flip-chip technology on a FR-4 board. To guarantee a small war-page over the test chip a low temperature process during curing is used. The chip has an under-fill to keep thermo-mechanical stress from the solder bumps and to stabilize the assembly during test. Figure 3 and 4 show a schematic of such a thermal test die with its implanted heat resistance and diode, the test chip assembly and the electrical interconnection.

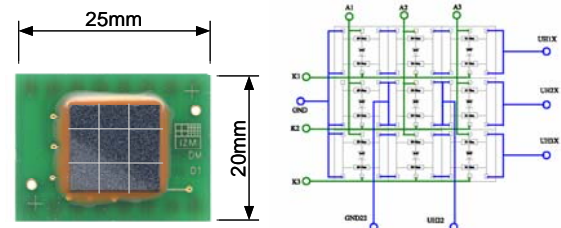

Fig. 3: Thermal test chip assembly and electrical interconnection schematic

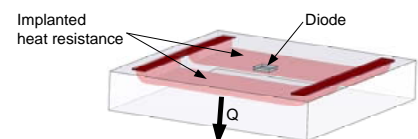

Fig. 4: Thermal test die structure

Determining the chip temperature $T_J$ the temperature-depending physical effect of the forward voltage, during a constant diode current, is used (equation 2).

$$T \sim U_{F,Diode} = \frac{KT}{e}\ln\left(\frac{I_{F,Diode}}{I_S}\right) @ I_{F,Diode} = const \quad (Eq.\ 2)$$

Using a linear fit through the measured voltage/temperature relations a numerical equation can be found to determine the junction temperature. For each thermal test chip its own characteristic has to be found.

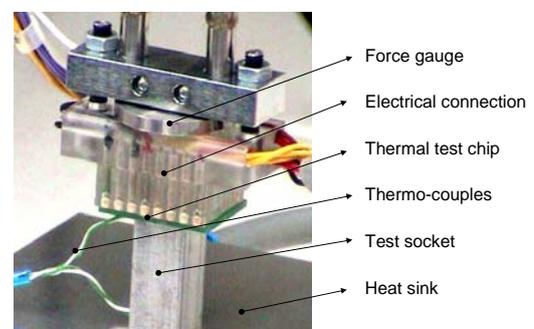

Fig. 5: New test set-up

The temperatures underneath the TIM surface in the test socket ($T_1$) and at the bottom of the socket ($T_2$) are measured by Ni-Cr thermo-couples (figure 5). Therefore holes are drilled into the socket close under the surface to the TIM and the bottom of the socket. The thermo-couples are assembled with heat conduction paste.
The measurement data generation is fully computer-controlled.

### 3.2. Principle

Usually the determination of the heat flow through a thermal interface material is assumed to be equal to the electrical power loss. But there is a parallel heat flow through the electrical connection block. Radiation and free convection could be neglected as shown in section 3.3. Using the presented new assembly structure the real heat flow through the thermal interface material can be determined. Figure 6 derives the thermal equivalent circuit for the set-up.

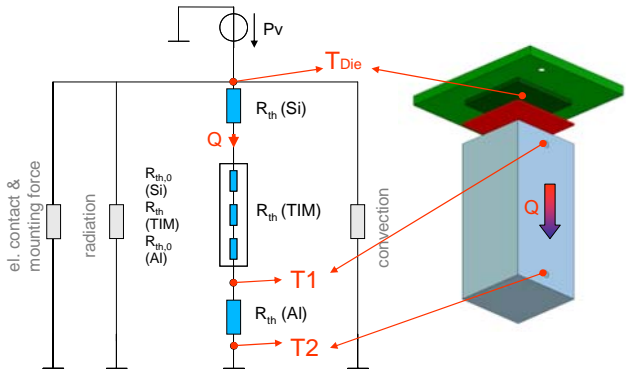

Fig. 6: Schematic of thermal equivalent circuit for test socket and test chip.

Measuring the temperature difference ($T_1$-$T_2$) in a physically known test socket the heat flow Q can be calculated (equation 3).

$$Q = \frac{T_1 - T_2}{R_{th,socket}} \quad ; \quad R_{th,socket} = \frac{d}{k \cdot A} \qquad (Eq.\ 3)$$

Knowing the heat flow through the test socket and the physical quantities of the thermal test chip, the effective thermal resistance $R_{th,eff}$ for the thermal interface material including the thermal interface resistance $R_{th,0}$ can be obtained (equation 4).

$$R_{th,eff} = \frac{T_J - T_1}{Q} - R_{th,Chip} \qquad (Eq.\ 4)$$

Having several measurements of the same TIM with different thickness the thermal conductivity and sum of both thermal interface resistances ($R_{th0,Si-TIM}$ and $R_{th0,TIM-Al}$) can be determined by numerical linear fit (figure 7).

$$R_{th,eff}(d) = \frac{d}{k_{TIM} \cdot A} + (R_{th0,1} + R_{th0,2}) = a \cdot d + b \qquad (Eq.\ 5)$$

Following the right part of equation 5, *a* describes the thermal conductivity normalized to the used area and *b* the thermal interface resistances (equation 6):

$$a = \frac{1}{k_{TIM} \cdot A} \quad , \quad b = (R_{th0,1} + R_{th0,2}) \qquad (Eq.\ 6)$$

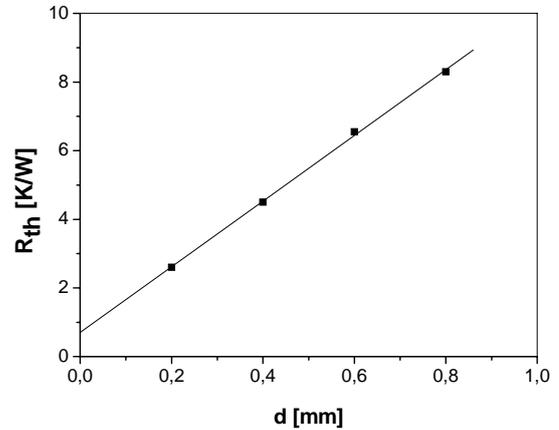

Fig. 7: Numerical linear fit through several measurements points of the same TIM with different thickness leads to the thermal conductivity of the TIM ($k_{TIM}$) and the sum of both thermal interface resistances ($R_{th0,Si-TIM}$ and $R_{th0,TIM-Al}$).

Knowing at least the values of the geometrical thickness and area, the thickness depending thermal conductivity of the thermal interface material can be obtained (equation 7).

$$k_{TIM}(d) = \frac{d}{(R_{th,eff}(d) - (R_{th0,1} + R_{th0,2})) \cdot A} \qquad (Eq\ 7)$$

### 3.3. Failure estimation and accuracy

To give a failure estimation and a accuracy for the set-up a thermal simulation was investigated and a evaluation of measuring failure was made [11].

The analysis has shown that the measurement error lies under 6%, when the temperature difference between $T_{Chip}$ and $T_1$ ($\Delta T_{Chip-1}$) is $\geq 5$ K.

For example, a TIM thickness of 50 µm results in a maximum measurable (effective) thermal conductivity of 10 W/mK. This is sufficient as most market-going adhesives do have values of $k_{eff\,(d=50\mu m)} \approx 0,5$-$1,5$ W/mK.

## 4. RESULTS
### 4.1. Measurements

Figure 8 shows the characteristics of four Ag-filled adhesives based on epoxy or modified epoxy bulk material.

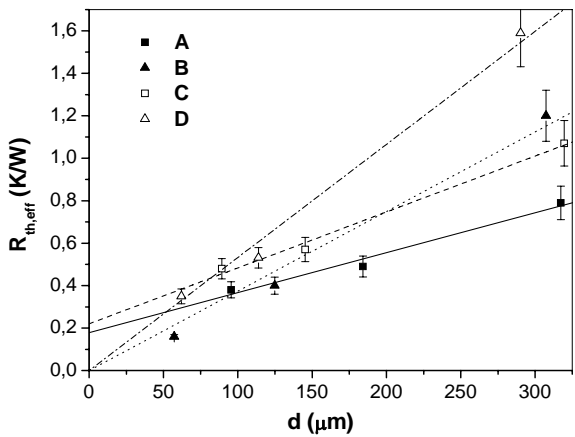

Fig. 8: Characteristic curves of four Ag- filled adhesives are used to determine its thermal conductivities and thermal interface resistances.

As derived above, the curves can be used to determine the thermal conductivities, the thermal interface resistances and the interface heat transfer coefficient respectively.

Therefore three specimens of different thickness for each adhesive were analyzed. After testing, cross-sections were made to measure the exact thickness of each specimen (figure 9).

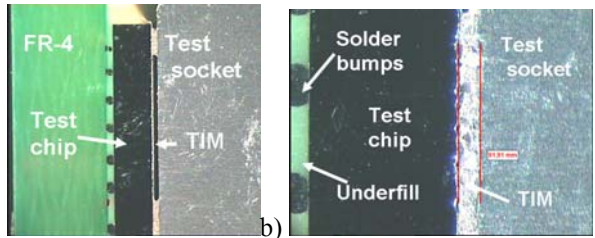

Fig. 9: Cross section of test chip assembled on test socket to determine the TIM thickness.

After numerical linear fitting, thermal conductivities and interface heat transfer coefficients could be determined (table 1).

Tab. 1: Determined parameters for investigated Ag filled Epoxy adhesives (A = 11,8 * 11,8 mm²).

| Investigated adhesive types | $k_{eff}$ (W/mK) (d = 100µm) | $k_{TIM}$ (W/mK) | $2*R_{th0}*A$ (K cm²/ W) |
|---|---|---|---|
| A | 1,9 | 3,8 | 0,24 |
| B | 2,5 | 1,9 | 0,07 |
| C | 1,4 | 2,7 | 0,31 |
| D | 1,3 | 1,3 | 0,014 |

**4.2. Metallurgy**

To get a better understanding of the structure of the TIM and its thermal interface resistance, the cross-sections were analyzed by SEM (figure 10).

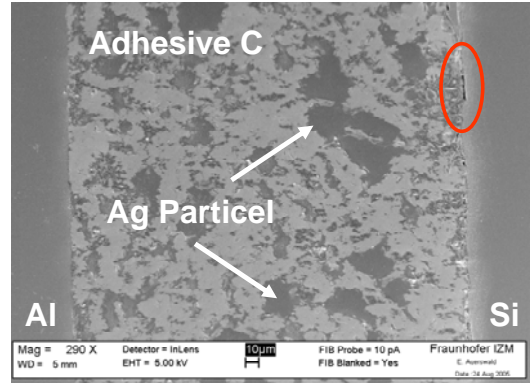

Fig. 10: Cross-section of TIM layer

Figure 11 shows a cut-out of figure 10 where in the boundary surface region pore structures would observed.

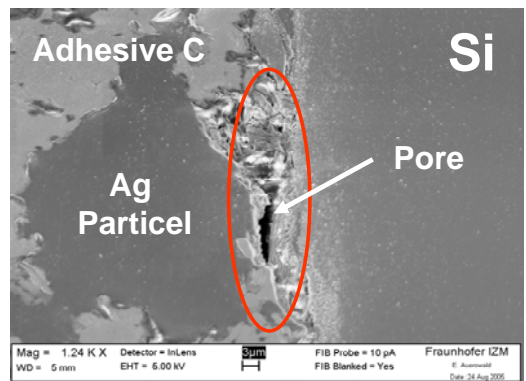

Fig. 11: Pore in the boundary surface region between the silicon chip and the Ag-filled copolymer TIM.

To have a better view into the depth of the interface, the boundary surface regions were analysed by focus ion beam (FIB) and SEM.

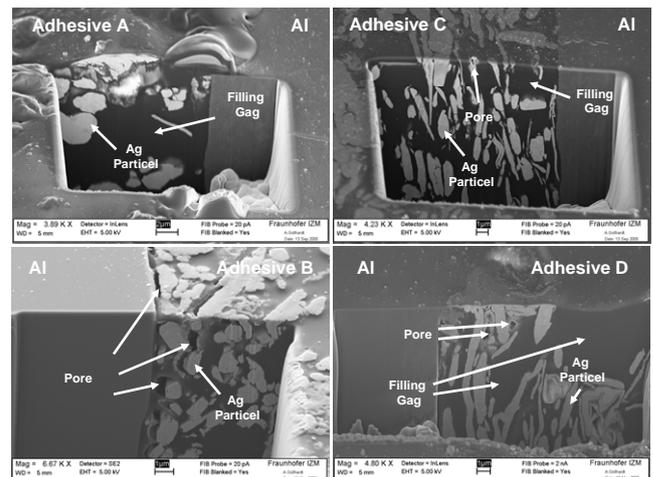

Fig. 12: Comparison of cross-sections for TIM-Si boundary section from Ag-filled adhesives .

Figure 12 compares cross-sections and FIB millings into the TIM-Al boundary sections of the four thermal interface materials.

It can be seen that there are different filling grades and that the Ag particles of adhesive A and C have a larger size then those of TIM B and D. Pores in the adhesive B could be observed mainly in the interface region. The other interfaces show a good connectivity. There also can be seen filling gaps between the Ag particles in the bulk material.

### 4.3. Simulation

To take into consideration of the influence of thermal interface resistance, a transient thermal simulation was done. Therefore the region between test chip surface and test socket was modeled using the effective conductivity $k_{eff}$ and using only the determined conductivity $k_{TIM}$ (without considering the thermal interface resistances) as given in table 1.

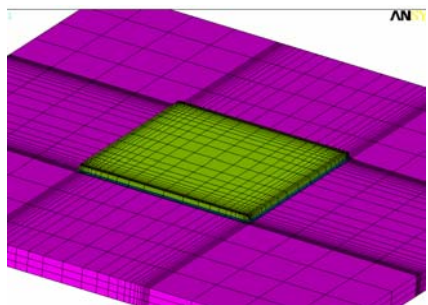

Fig. 13: Model for transient simulation.

Figure 13 shows the model structure. The chip is assembled on a FR-4 PCB. The die attach area is filled with thermal via's. Table 2 gives the model parameters.

Tab. 2: Model parameters for investigated transient thermal simulation.

| $A_{Chip}$ [mm²] | $d_{Chip}$ [μm] | $d_{TIM}$ [μm] | $d_{PCB}$ [μm] | $d_{CU}$ [μm] | th. via count | $dia_{Via}$ [μm] |
|---|---|---|---|---|---|---|
| 8 x 8 | 360 | 100 | 1600 | 70 | 100 | 500 |

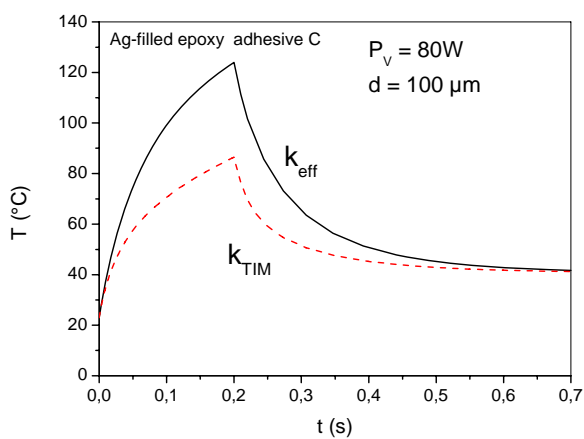

Fig. 14: Transient temperature behaviour in respect with and without considering the thermal interface resistances.

Figure 14 compares the transient behavior. As expected, the observed temperature rise for the solution of the model, considering only the thermal conductivity $k_{TIM}$, is much less than considering the effective conductivity.

That means that in real systems the junction temperature will be much higher as through simulation expected. This caused an over heating of the device and in the end its damage.

## 5. DISCUSSION OF RESULTS

The metallurgical results have shown that it is necessary to observe the TIM and its boundary regions. The Ag particles of adhesives B and D have for instance a closer contact between each other as the Ag particles of TIM A or C which results in a higher thermal bulk conductivity (c.f. figure 8). The observed filling gaps and pores in the bulk material could be a reason for the difference of thermal interface resistances for the investigated four adhesives. Especially if pores are close to the interface region, as seen in figure 11, the thermal interface resistance will be influenced.

This is in correspondence with the measured results. TIM A and C have larger Ag particles as B and D and they show a better thermal conductivity because of its larger particle surface, which leads to less interfaces between the particles over the TIM thickness d thus enhancing heat transfer.

On the other hand larger particles have an influence on the thermal interface resistance as they seem to promote the development of pores in the interface region and gaps in the bulk material. Opposed to that, TIM B has small Ag particles which shows a good filling grade even in the interface region. In this case the thermal conductivity decreases (due to a higher number of interfaces between the filler particles) but the thermal interface resistance becomes negligible. TIM D have very flat and long Ag particles which show a good accommodation to the interface and therefore a low thermal interface resistance. Because of the oblong Ag particles the number of interfaces between the Ag particles in the polymer matrix increases and for that the thermal conductivity decreases, too.

Equations 10 and 11 describe the behaviour of effective thermal conductivity $k_{eff}$ in respect to TIM thickness d. It can be concluded that for thin TIM layers the effective thermal conductivity is a function of both the thermal interface resistances $R_{th0,1} + R_{th0,1}$ and the bulk conductivity $k_{TIM}$.

$$k_{eff}(d) = \frac{d}{\frac{d}{k_{TIM}} + (R_{th0,1} + R_{th0,2}) \cdot A} \quad (Eq\ 10)$$

For thick TIM layers the effective thermal conductivity follows from the bulk resistance.

$$k_{eff}(d \to \infty) = k_{TIM} \qquad (Eq\ 11)$$

Figure 15 shows for the investigated TIMs the mentioned behaviour.

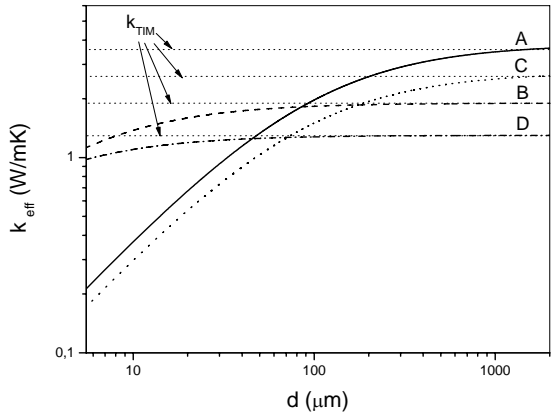

Fig. 15: Behavior of effective thermal conductivity in respect to the thickness for the investigated Ag-filled adhesives.

It can be recognized that the adhesive A has for thicker layers a better effective thermal conductivity as C,B and D. But having a thermal design using a thin TIM layers e.g. the investigated adhesive B or D should be taken.

The transient thermal simulation has shown that it is necessary to take the thermal interface resistance into consideration. Otherwise it causes overheating of the device. Even after 200ms the temperature for the model that considers $k_{eff}$ rises up to more than 50% of the temperature of the $k_{TIM}$ based model.

## 6. CONCLUSION

A new test set-up to characterize thermal interface materials was investigated.

The novelty is that the set-up allows to measure all classes of TIMs (incl. solder and adhesive) under real assembly conditions.

The advantages are:
- High versatility (all TIM).
- A rapid test method using a test socket.
- An inexpensive method using standardized reference test socket material with narrow tolerances.
- Obtaining all relevant of information concerning: thickness, force, $k_{eff}$, $k_{TIM}$, $R_{th0,i}$
- TIM thickness determination by cross-sectioning.
- Design allows easy analysis by FIB and/or SEM
- Flip-chip assembly for thermal test chip used as to assure flatness of die (important for greases)
- Test uses surfaces as in real devices
- More accuracy is possible to improve temperature measurement while using thermocouples with less measurement errors.

The determination of thermal interface resistance and thermal conductivity was exemplified for four adhesives. The Structure of the adhesive was analyzed by FIB and SEM. It was found, that there is a structure-property correlation with respect to thermal properties. Voids, particle size and filler content could be shown to influence interface resistance and bulk conductivity and was explained accordingly.

Also the influence of the thermal interface resistance was shown by thermal simulation.

## 7. REFERENCES

[1] S.V. Garimella, Y.K. Joshi, A. Bar-Cohen, R. Mahajan, K.C. Toh, V.P. Baelmans, J. Lohan, B. Sammakia, and F. Andros. 'Thermal challenges in next generation electronic systems – summary of panel presentations and discussions' IEEE Trans. On Components and Packaging Technologies, 25(4):569–575, 2002.

[2] R. Viswanath, V. Wakharkar, A. Watwe and V. Lebonheur. 'Thermal Performance Requirements from Silicon to Systems', Intel Technology Journal Q3, pp. 1-16, 2000.

[3] Blazej, D. 'Thermal Interface materials', Electronics Cooling, Volume 9, Number 3, August 2003, S. 14 ff

[4] Bruce M. 'Calculations for thermal interface materials';, Electronics Cooling, Volume 9, Number 3, August 2003, S. 8 ff

[5] Lasance, C. J.M. 'Problems with Thermal Interface Material Measurements: Suggestions for Improvement', http://www.electronics-cooling.com/html/2003_november_a2.html

[6] Bosch, E., Lasance C. 'High accuracy thermal interface resistance measurement using a transient method' Electronics Cooling, Vol.6, No.3, September 2000

[7] Netzsch TCT416 , http://www.netzsch.de

[8] http://www.thermagon.com/pdf/Resistance.pdf

[9] Wöstmann, F.-J, Hahn, O. 'Messverfahren zur Ermittlung der Wärmeleitfähigkeit dünner Klebschichten', Produktion von Leiterplatten und Systemen, Heft 1, S. 139ff, 2005

[10] May, D. 'Entwurf und Aufbau eines Messplatzes zur Charakterisierung von Thermischen Interface Materialien', Diploma Thesis, FHTW, Berlin, Germany, 2004

[11] Schacht, R., May, D., Wunderle, B., Gollhardt, A., Wittler, O., Michel, B., Reichl, H. 'Characterization of Thermal Interface Materials for Thermal Simulation', ESTC 06, Dresden, Germany, 2006